# Verification-Time Dependency on a Disappearing Evaluator

### An Operational Protocol for Decision-State Commitment, Independent Verifiability, and Stability-Calibrated Counterfactual Auditability

**Ho Wa KU | Jameel Ahmed Siddiqui**

## Abstract

AI governance and assurance frequently assume that a consequential model-mediated decision can be reconstructed or tested after the fact. That assumption may fail when the evaluator that produced the decision is no longer accessible in the same version and execution context. Commercial model lifecycles and regulatory record-retention duties are different clocks: Article 18 of the EU AI Act requires specified provider documentation for high-risk AI systems to be kept for ten years, while Article 19 separately requires automatically generated logs under provider control to be retained for at least six months, subject to applicable law. Neither provision guarantees that the original evaluator remains callable or re-instantiable. This paper derives three verification-time constructs from the published Execution Governance (EG) 3.0 architecture: Decision-State Commitment, Independent Verifiability, and Counterfactual Auditability. Independent reprocessing of the released Study 2 artifacts reproduces the original within-family behavioural comparisons: 52.0% modal-decision reversal for Llama 3.1 8B versus Llama 3.3 70B (26/50; 95% Wilson 38.5-65.2) and 30.0% for GPT-OSS 20B versus GPT-OSS 120B (15/50; 19.1-43.8). The Retiring Witness v1.1 corrects the earlier predecessor-successor label: those were within-family comparisons, not provider-established succession. Groq's published migration map instead designates Llama 3.1 8B -> GPT-OSS 20B and Llama 3.3 70B -> GPT-OSS 120B (or Qwen for the latter). Post-hoc re-pairing of the already released outputs against the two GPT-OSS replacement paths yields 64.0% and 38.0% reversal, respectively. These replacement-path figures are descriptive reanalyses, not pre-specified controlled experiments: the Llama runs used max_tokens=120 without a reasoning-effort setting, whereas GPT-OSS used max_tokens=900 with reasoning_effort=low. The 38% path is therefore treated as the stronger non-degenerate replacement-path observation; the 64% path remains a boundary observation because the Llama 3.1 model approved 49/50 cases. Within each original family, 2 of 100 model-case pairs exhibited decision variation across three temperature-zero repetitions, and exact generated-reason text varied more often than the decision label under the stated exact-text criterion. The joint contribution is an operational verification-time protocol and an optional Verification-Time Preservation Package (VTPP) specifying what a proposed higher-assurance profile binds at authorization time, what a separately trusted verifier can substantiate later, which semantic and cross-field conformance checks sit beyond schema validity, how stability and paired counterfactual tests should be calibrated, and what evidence should be preserved when later access to the original evaluator cannot be assumed. The protocol is downstream and non-authorizing: it does not add a seventh EG live condition, alter the EG Core Formula, or state jurisdiction-specific legal admissibility.

## 1. Introduction

Post-hoc review of an AI-mediated decision can inspect at least three different objects: the record of the decision, the system or evaluator that produced it, and the behavior of that evaluator under alternative inputs. These are not interchangeable. A durable record can preserve what happened even after a hosted model version is retired, but it cannot by itself reproduce how the model would respond under a later counterfactual query.

The verification problem is therefore temporal as well as epistemic. An evaluator may be operationally unavailable to the reviewer even though documentation about the system remains subject to long retention periods. Google Cloud publishes model lifecycle stages, retirement dates, migration guidance, and retired-model lists. For open or partner models offered through managed Model-as-a-Service surfaces, Google publishes separate shutdown schedules. Anthropic likewise publishes retirement schedules for Anthropic-operated platforms and notes that partner-platform schedules can differ. The re-verified census deposited with The Retiring Witness v1.1 makes this surface dependence concrete: Claude Sonnet 3.7 and Claude Haiku 3.5 ceased availability on Anthropic-operated services on 19 February 2026, while the corresponding Google Cloud/Vertex surfaces had later shutdown dates. Evaluator identity for later verification therefore cannot safely be reduced to a model-version string alone; the relevant hosting/service surface and time are part of the availability state. These sources establish the operational possibility that later managed-service/API access to the same evaluator can end at different times on different surfaces. They do not imply that model weights have been destroyed, that every provider follows the same lifecycle, or that no escrow, self-hosting, preservation, or research-access arrangement is possible.

The research question is narrower than "Can the old model be recovered?" It is: what evidence and test procedure must exist so that a later reviewer can distinguish (i) a preserved account of what was authorized, (ii) a verifiable account of the basis on which it was authorized, and (iii) a statistically defensible claim about what the evaluator would have done under a specified alternative?

## 2. Prior Work and Contribution Boundaries

This paper follows the provenance discipline used in the authors' earlier joint technical note, Before the Effect and After the Act (DOI: 10.5281/zenodo.21201653): prior work remains attributed to its origin, and only the new procedure developed through the present exchange is claimed jointly.

Architectural baseline (KU). The prior EG architecture used in this paper is limited to canonical published semantics: the six-condition Core Formula; Authorization Continuity; RecordBound and CommitCoupled; the Effect Authorization Receipt (EAR) or equivalent final authorization envelope; the Governed Effect Record; Reviewability; Sufficient Verifiable Proof; verifier-independence declarations; Material Evolution / Compatible Continuation; and capability-authority separation. Those semantics are prior work by KU and are not newly claimed here.

Derived verification-time constructs. The labels Decision-State Commitment, Independent Verifiability, and Counterfactual Auditability are used in this paper as verification-time specializations derived from the canonical EG 3.0 semantics above. They are not represented here as additional EG Core Formula primitives or as a seventh live condition. This distinction is important for terminology and provenance.

Empirical baseline (Siddiqui). The retirement-event census, the original within-family behavioural comparisons, repeated-run instability measurements, instruments, and released raw data are prior work by Siddiqui, now versioned as The Retiring Witness v1.1 (DOI: 10.5281/zenodo.22122318). Version 1.1 explicitly corrects the earlier predecessor-successor relationship label while leaving the measurements and original artifacts unchanged. This paper does not relabel those measurements as joint work. KU independently reprocessed the released Study 2 files and additionally performed the provider-designated replacement-path reanalysis from the same raw outputs after checking Groq's published migration map. That reanalysis does not change the provenance of Siddiqui's underlying data.

Joint contribution. The jointly claimed contribution is the operational protocol in Sections 4-6: the explicit EG 3.0 crosswalk, a proposed verification-time assurance predicate, a preservation and testing procedure, a claim-evidence taxonomy, stability-calibrated counterfactual auditability, VTPP, and the interpretation of what the corrected empirical taxonomy means for verification-time assurance. The provider-designated replacement-path calculations are an additional analysis by KU using Siddiqui's released responses; their methodological treatment and claim boundaries are jointly agreed in this manuscript.

## 3. Verification-Time Dependency and the Empirical Baseline

### 3.1 Regulatory retention is not evaluator retention

The draft v1 treated the EU AI Act as if it created a single 120-month "documentation and record-keeping horizon" for high-risk AI systems. That is too broad. Article 18 requires providers, for ten years after a high-risk AI system is placed on the market or put into service, to keep specified documentation available to national competent authorities: the Article 11 technical documentation, Article 17 quality-management-system documentation, specified notified-body documents where applicable, and the Article 47 EU declaration of conformity. Article 19 separately requires automatically generated logs, to the extent under provider control, to be kept for a period appropriate to the intended purpose and at least six months unless applicable Union or national law provides otherwise. Neither provision states that the exact model version, model weights, or execution endpoint must remain callable for ten years. The EU rules are therefore used here only to illustrate a potential mismatch between documentary retention and evaluator availability.

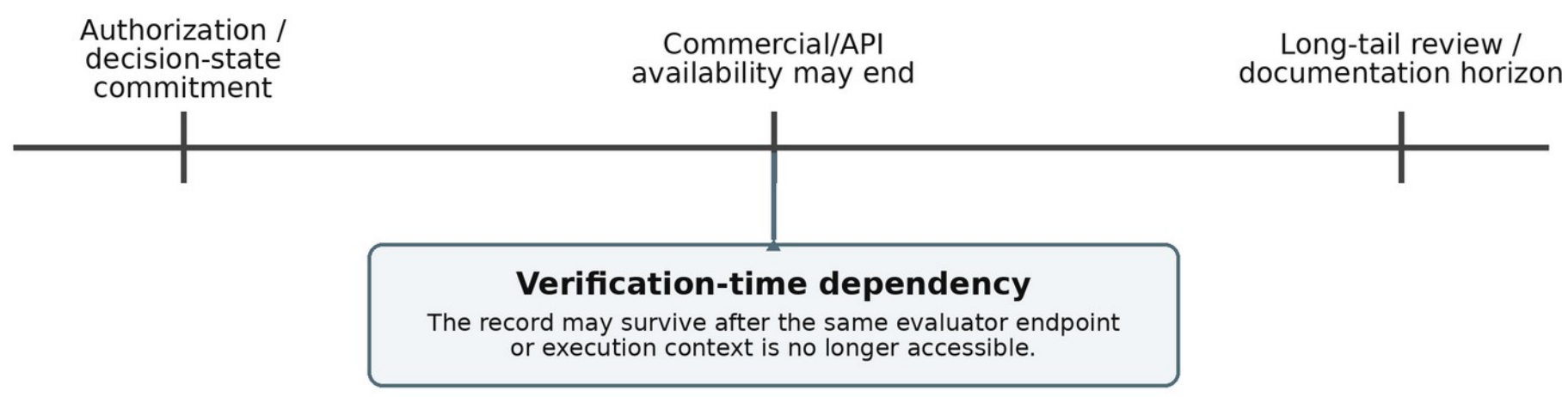

*Figure 1. Verification-time dependency: commercial availability and documentary retention are different clocks.*

### 3.2 Reported retirement-event census

Siddiqui's prior publication reports 22 model retirement events across three providers and a median reported lifespan interval of 16.4 months under its published release-to-retirement method. Independently recomputing the frozen v1.1 CSV yields a median of 16.45 months, a mean of 18.72 months, a range of 3.9-40.3 months, and 17/22 intervals below 24 months. These figures reproduce the published baseline, but they should not be read as a channel-specific continuous-availability distribution: the v1.1 verification pass establishes retirement-date provenance, while the historical release_date field is retained from the published census and is not redefined here as the first callable date on every service surface.

Version 1.1 (DOI: 10.5281/zenodo.22122318) deposits retirement_census_v1_1_verified.csv and README_retirement_census_v1_1.txt. On 28 August 2026, the exact frozen CSV supplied directly by the coauthor was independently hashed: SHA-256 b63a15f5d87ae87b7b9b5afc0fced7acac8d50e344ebd2e998fa76a76ff9d324 and MD5 9a14876a3e48a5283cfffdc190f5f329. The SHA-256 exactly matches the digest stated in the frozen

README, and the MD5 matches the repository value recorded for the census. The directly supplied README itself has SHA-256 4a051d29213aa83c9136e627b841c65027125a17e4393cdb3957b0f47b2230a2. All 22 lifespan_days and one-decimal lifespan_months values recompute exactly from the frozen dates using the stated 30.44-day-month method.

Current provider documentation continues to support the material retirement-date and service-surface qualifications. Anthropic-operated services retired Claude Sonnet 3.7 and Claude Haiku 3.5 on 19 February 2026, while Google Cloud retained later shutdown dates for the corresponding partner-model surfaces; Claude Haiku 3 likewise had a later Google Cloud shutdown. For Gemini 2.5 Pro and Flash, Google's 2 April 2026 release notes recorded 16 October 2026, while the current lifecycle table lists 20 October 2026. The frozen v1.1 census therefore retains 16 October only as a historical planned-date observation flagged soft_no_earlier_than; this paper does not treat 16 October as the current fixed retirement date. The census remains a 22-event sample across three providers, not a claim to enumerate every retirement event.

### 3.3 Original within-family behavioural comparisons

Independent reprocessing reproduces the two comparisons actually specified by the released runner. Family I compares llama-3.1-8b-instant with llama-3.3-70b-versatile and reverses 26/50 modal decisions, 52.0% (95% Wilson 38.5-65.2). The first model approves 49/50 cases (98%) and the second 23/50 (46%); all 26 reversals are APPROVE-to-DECLINE. Because only one first-model DECLINE existed, at most one DECLINE-to-APPROVE reversal was available. The 52% result is therefore retained as a near-degenerate within-family boundary comparison, not as evidence of provider-established succession.

Family II compares openai/gpt-oss-20b with openai/gpt-oss-120b and reverses 15/50 modal decisions, 30.0% (95% Wilson 19.1-43.8). Approval rates are 17/50 (34%) and 6/50 (12%), with 13 APPROVE-to-DECLINE and 2 DECLINE-to-APPROVE reversals. OpenAI released gpt-oss-20b and gpt-oss-120b together on 5 August 2025 as two differently sized open-weight models; this comparison is therefore a within-family/model-scale behavioural comparison, not a predecessor-successor sequence.

Across the two original within-family comparisons, only four cases reverse in both and 37/50 cases (74%) reverse in at least one comparison. The two first-listed models agree on 18/50 cases (36%) and the two second-listed models on 31/50 (62%). These descriptive cross-family summaries remain reproducible from the released data but should not be read as evidence of a common succession process or population-level replacement rate.

### 3.4 Provider-designated replacement-path reanalysis (post hoc)

Groq's deprecation history documents two relevant recommended migration paths on the same host: llama-3.1-8b-instant -> openai/gpt-oss-20b, and llama-3.3-70b-versatile -> openai/gpt-oss-120b (with qwen/qwen3.6-27b also offered for the latter). Because all four models had already been run on the same 50 cases, the released outputs can be re-paired against those provider-designated paths without issuing new model calls. This is a post-hoc reanalysis, not a pre-specified replacement-path experiment. A further asymmetry must remain explicit: the Llama runs used max_tokens=120 and no reasoning_effort parameter, whereas the GPT-OSS runs used max_tokens=900 and reasoning_effort=low. The same cases, host, decision format, system instruction and temperature do not remove that cross-family invocation difference.

The stronger non-degenerate replacement-path observation is llama-3.3-70b-versatile -> openai/gpt-oss-120b: 19/50 modal decisions reverse, 38.0% (95% Wilson 25.9-51.8), with approval rates 23/50 (46%) -> 6/50 (12%) and 18 APPROVE-to-DECLINE versus 1 DECLINE-to-APPROVE reversal. This supports only a narrow substitution warning: on this provider-designated path and instrument, the replacement output cannot be presumed behaviourally equivalent to the retired evaluator.

The second provider-designated path, llama-3.1-8b-instant -> openai/gpt-oss-20b, reverses 32/50 modal decisions, 64.0% (95% Wilson 50.1-75.9), with approval rates 49/50 (98%) -> 17/50 (34%) and all 32 reversals APPROVE-to-DECLINE. It is retained as a boundary replacement-path observation because the 98% starting approval rate mechanically constrains reversal direction.

Using the same token-set Jaccard definition implemented in run.php, post-hoc mean generated-reason overlap is 0.2783 for Llama 3.1 8B -> GPT-OSS 20B and 0.2610 for Llama 3.3 70B -> GPT-OSS 120B. These values are descriptive lexical-overlap measures only. They are not semantic-equivalence tests, and the cross-family invocation asymmetry applies to their interpretation as well.

| Analysis class | Pair and observed reversal | Interpretive status |
|---|---|---|
| Original within-family | Llama 3.1 8B vs Llama 3.3 70B: 26/50 (52%) | Boundary behavioural comparison; 98% first-model approval rate; not provider succession. |
| Original within-family | GPT-OSS 20B vs GPT-OSS 120B: 15/50 (30%) | Within-family/model-scale comparison; same-day sibling release; not provider succession. |
| Provider-designated post-hoc | Llama 3.3 70B -> GPT-OSS 120B: 19/50 (38%) | Stronger non-degenerate replacement-path observation; post-hoc and cross-family invocation asymmetry disclosed. |
| Provider-designated post-hoc | Llama 3.1 8B -> GPT-OSS 20B: 32/50 (64%) | Boundary replacement-path observation; 98% starting approval rate; post-hoc and parameter-asymmetric. |

*Table 1. Empirical relationship taxonomy after baseline correction and provider-migration reanalysis.*

### 3.5 Within-version decision instability

Each released family contains 100 model-case pairs (50 cases x 2 models) with three repetitions per pair, or 300 responses per family. Independent reprocessing finds 2 response departures from the pair-modal decision in the Llama family and 2 in the GPT-OSS family; in each family those departures occur in 2 of 100 model-case pairs. Thus the descriptive response-level rate is 2/300 = 0.67% per family, while the pair-level incidence is 2/100 = 2.0% per family. A simple-binomial Wilson interval for 2/300 is numerically 0.18-2.40%, but the independence assumption required for inferential use is not established because responses are nested within heterogeneous model-case pairs and share hosted-backend conditions. These values are therefore descriptive stability checks, not universal inferential noise floors.

### 3.6 Reason-channel instability

Independent reprocessing corrects an important denominator/conditioning detail. In the Llama family, exact generated-reason text varies in 40 of 100 model-case pairs; 38 of those 100 pairs retain a unanimous decision across all three repetitions while the reason text changes. In the GPT-OSS family, exact generated-reason text varies in 24 of 100 pairs, but only 22 of 100 combine a unanimous decision with changing reason text; the other two are the decision-unstable pairs. The earlier statement that all 24 GPT-OSS reason-varying pairs had unanimous decisions is therefore corrected. Using the token-set Jaccard function implemented in run.php over each model's concatenated reasons, the independently reprocessed mean within-family paired-model reason overlap is 0.257 for Family I and 0.342 for Family II. These are tokenizer/normalization-specific lexical measures, not semantic equivalence tests. No inferential 36-fold comparison is claimed, and these findings do not apply to deterministic EG authorization-controller reason_codes unless such codes are themselves model-generated and tested.

| **Claim from draft v1** | **Status after fact-check** | Treatment in this paper |
|---|---|---|
| EU AI Act = 120-month documentation and record-keeping horizon | Overbroad | Corrected: Article 18 = specified documentation for 10 years; Article 19 = logs at least 6 months where under provider control. |
| 15/50 GPT-OSS within-family reversals = 30.0% [19.1, 43.8] | Arithmetic verified; relationship label corrected | Retained as an original within-family/model-scale behavioural comparison, not successor substitution. |
| 2/300 non-modal responses = 0.67% [0.18, 2.40] | Arithmetic verified; inferential use requires clustering assumptions | Retained as a descriptive response-level rate; simple Wilson interval labelled descriptive because responses are clustered within model-case pairs. |
| 24% GPT-OSS reason variation is ~36x decision instability | Invalid ratio and earlier unanimity wording overstated | Corrected: 24/100 any exact reason variation; 22/100 unanimous-decision reason variation; no 36x inference. |
| 30 repetitions is an adequate universal floor | Not supported by Wilson precision | Removed; repetition count is determined from a pre-specified precision/power target. |
| 52% / 30% within-family reversals and reason Jaccard 0.257 / 0.342 | Independently reprocessed; original successor label unsupported | Retained as original within-family comparisons. Provider-designated replacement paths are analysed separately and explicitly post hoc. |
| Provider-designated replacement-path reanalysis: 64% / 38% | Independently recomputed from released outputs; not pre-specified; cross-family invocation parameters differ | Retained as post-hoc descriptive reanalysis. 38% is the stronger non-degenerate observation; 64% is a 98%-baseline boundary case. |

### 3.7 Independent raw-file reprocessing and evidence identity

The original released package contains cases.json, results_gptoss.csv, results_llama.csv, run.php, and The_Retiring_Witness_Siddiqui_Year-2026.pdf. Independent reprocessing against Zenodo record 21737306 matched the repository-displayed MD5 values and computed SHA-256 as follows: cases.json cd9eeaaf22b50a6bfdf491283c7489fa76f5f4dd3e1b83e0557896d5c83748d8; results_gptoss.csv 175473266d464eab43c20cae7366d8308441ae1bef5b21fb7455f0226e38061b; results_llama.csv 9814449080eb8171af54b64b2bebf151c671ad1f9f9aa38101e8d0a5c09505f0; run.php 37e38df4374efd5e15237616e07898584a3d713075ba302d8c683531eb3c05a3; and the PDF 92147dbbedd028ea3368d1f973a4b89d3e1a09bd104d17229692cffe092a3d33. Zenodo version 1.1 (22122318) publicly lists the same MD5 values for these original artifacts, consistent with the coauthor's statement that they are byte-identical, while adding the erratum and census files. The v1.1 copies themselves were not re-downloaded for a second byte-level comparison in this review. The replacement-path reanalysis uses the already verified local raw files and does not require new model calls.

The companion R1 Independent Reprocessing Manifest records the original family-level reversals, within-model variation, exact-text reason variation, Jaccard values under the released run.php tokenization, cross-family overlap/agreement, and file hashes. Version 8.3 additionally records the provider-designated replacement-path calculations as post-hoc reanalysis, with the original runner's max_tokens/reasoning_effort asymmetry treated as a mandatory limitation. Original Study 2 raw-file reprocessing is closed for the retained measurements. Exact-byte verification of the directly supplied frozen v1.1 census CSV is also closed: its independently recomputed SHA-256 and MD5 match the stated README digest and recorded repository MD5. This byte-identity result does not enlarge the census's coverage or release-date provenance claims.

## 4. The Three-Layer Verification Chain

The core architectural insight survives the fact-check, but the terminology is now anchored explicitly to canonical EG 3.0 semantics rather than treating the three verification-time labels as pre-existing EG primitives.

### 4.1 Canonical EG 3.0 Crosswalk

| Joint-paper construct | Canonical EG 3.0 source semantics | Status in this paper |
|---|---|---|
| Decision-State Commitment | RecordBound; EAR/equivalent final authorization record; Governed Effect Record; input/context/evidence and commitment metadata | Derived verification-time specialization: preserves the state materially relied upon at authorization/commitment. |
| Independent Verifiability | Sufficient Verifiable Proof; verifier competence/trust; verifier-independence declaration; EBL-5 where independent assurance is actually demonstrated | Higher-assurance specialization. An independent claim requires attributable verifier participation evidence and trust/control separation; shared administrative control is a non-independent claim, not a lower grade of independence. |
| Counterfactual Auditability | Reviewability; evidence retention/provenance; Material Evolution / Compatible Continuation; evaluator/version identity | New downstream verification-time research construct. It is not a Core Formula condition and does not create authority. |
| Verification-Time Preservation Package (VTPP) | Profile-defined assurance evidence; EAR/Governed Effect Record evidence references; retention/disclosure responsibilities | Optional higher-assurance evidence profile. It is subordinate to the Core Formula and cannot turn non-authority into authority. |
| Model-generated explanation channel | Distinct from controller-produced EAR reason-code / deterministic authorization-controller code | Empirical instability claim applies to model-generated explanation text unless a controller reason-code is itself model-generated and tested. |

### 4.2 Proposed Verification-Time Assurance Predicate

EG 3.0 answers the commitment-time authorization question through the existing MayCommit / CommitEffect invariant. This paper introduces a separate, non-normative verification-time assurance predicate for a later review time t_v. It is deliberately downstream of EG authorization and does not modify the Core Formula:

```
VTA(q,r,m,t_v) := VerifyRecord(r,t_v) AND [RecordSufficient(q,r) OR PreservedCounterfactualEvidence(q,r)
OR (EvaluatorReinstantiable(m,t_v) AND PreSpecifiedProtocol(q))]
```

The predicate means only that a declared verification claim q has a preserved basis at review time. RecordSufficient(q,r) means that the preserved record contains the evidence required for that declared claim class under a pre-specified verification rule; a recorded assertion does not become true merely because it was recorded. The predicate does not establish that q is true, that the original authorization was lawful, or that a later counterfactual is available. In particular, current authority at commitment does not imply later behavioral verifiability: Authorized(t_c) does not by itself imply Verifiable(t_v).

### 4.3 Decision-State Commitment

Decision-State Commitment is the paper's verification-time specialization of EG record binding and durable effect-record semantics. It preserves the authorization-time state materially relied upon. At minimum, the commitment should identify the evaluator/version or applicable identity epoch, input or input hash, relevant system/developer instructions where available, mandate or policy version, constraints, evidence/context snapshot, tool permissions, timestamp, authorization outcome, and exact target effect. It answers: what state was bound to the authorization decision and later commitment?

### 4.4 Independent Verifiability

Independent Verifiability is used here only for claims that seek an independent assurance label. For such a claim, a verifier must be separately trusted from the original decision producer/operator and must be able to validate integrity and provenance without relying solely on that producer's assertion. The assurance claim must itself be evidenced: the package must identify the decision producer and package author, identify the verifier's administrative/signing domain or key, and carry a verifier-generated signature or attestation reference showing that the declared verifier actually handled the package or verification result. A recorded statement that verification was independent is not sufficient. Where the verifier shares administrative control with the decision producer or package author, the assurance is weaker in kind, not merely lower in degree: it is a non-independent or common-control verification claim and must be labelled as such. Separate legal incorporation is not the sole criterion, but administrative/control separation, signing/attestation provenance, competence, custody, and common-mode failure risks must be explicit. Where independence is not claimed, the relationship and residual limitation must be disclosed. As an adjacent role-separation precedent, IETF RFC 9334 (RATS) distinguishes an Attester that produces Evidence, a Verifier that appraises Evidence, and a Relying Party that consumes an Attestation Result. This paper cites that separation only as related assurance architecture; VTPP does not claim RATS conformance, interoperability, endorsement, or identity between those roles and EG/VTPP roles.

### 4.5 Counterfactual Auditability

Counterfactual Auditability addresses a different question: under a specified perturbation, would the evaluator's output have changed beyond its own measured instability? This requires an executable evaluator or contemporaneously preserved counterfactual evidence. A record of the original decision cannot answer an unknown future counterfactual by itself.

### 4.6 Evaluator persistence is a precondition for open-ended future counterfactuals

This creates a hard boundary. If a later dispute raises a counterfactual that was not pre-specified and the exact evaluator cannot be re-instantiated, the system may be unable to answer that question empirically. The appropriate governance response is not to substitute a replacement model silently. It is to state the limitation and distinguish preserved decision evidence from unavailable behavioral evidence.

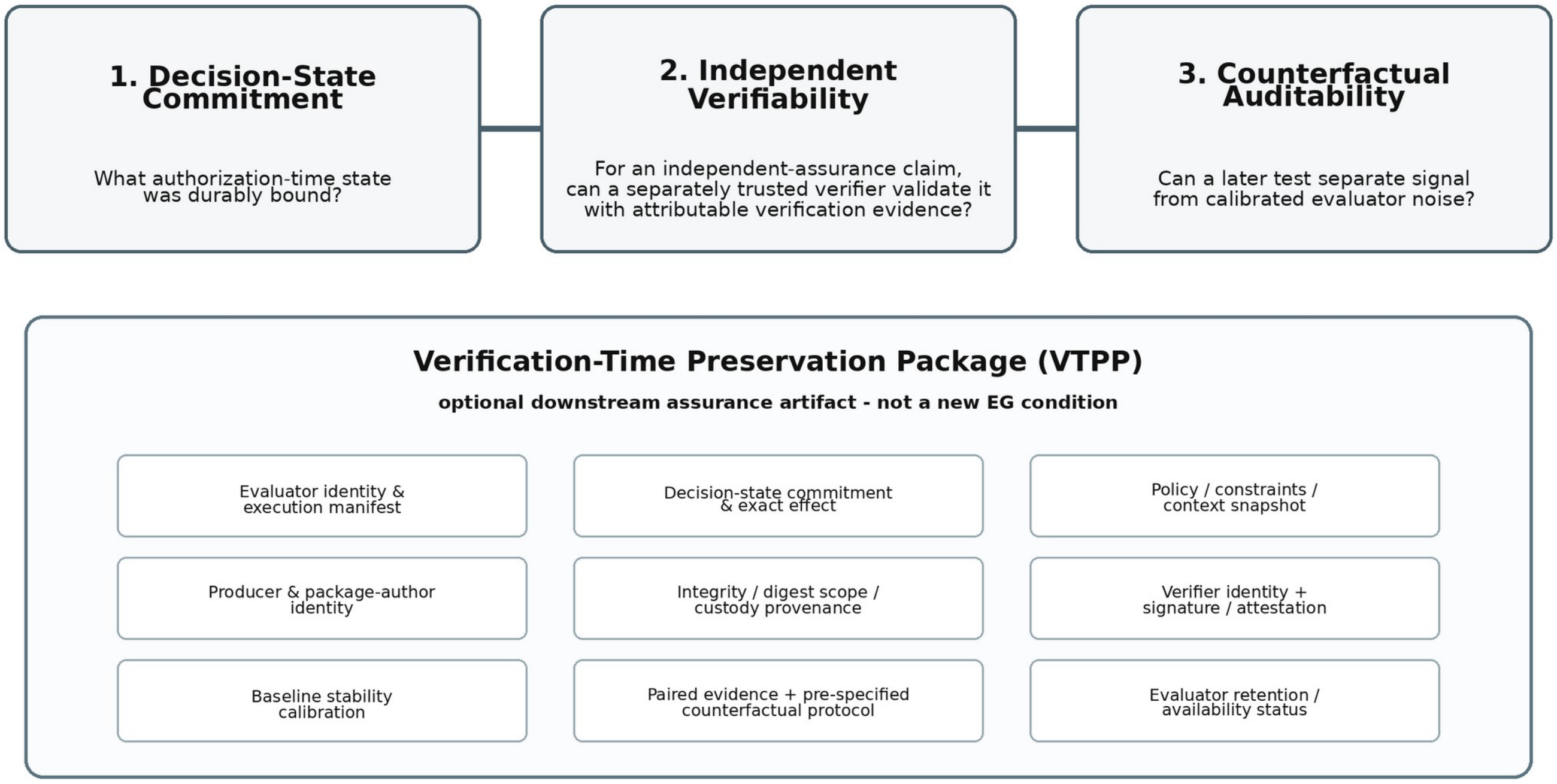

*Figure 2. Three-layer verification chain and the proposed Verification-Time Preservation Package.*

## 5. Operational Verification-Time Preservation Protocol

This section is the principal joint contribution. It is deliberately framed as an audit and research protocol. Terms such as "admit" and "exclude" refer to the protocol's own evidentiary gate, not to the law of evidence in any particular jurisdiction.

### 5.1 Scope and unit of analysis

The unit of analysis is a model-mediated authorization event: an event in which a model output materially contributes to whether a consequential action is permitted, escalated, held, or blocked. The protocol does not assume that the model is the sole decision-maker. Human review, deterministic policy, and external evidence may all be part of the authorization state and should be committed if relied upon.

### 5.2 Preconditions

1. A Decision-State Commitment exists and is created contemporaneously with the authorization process, with its state binding fixed no later than finalization of the authorization decision. If a governed effect is later committed, the required authorization-state binding must exist before that effect commitment. It identifies the evaluator, relevant input state, policy/mandate and constraints, evidence/context relied upon, and target effect.
2. Where an independent verification-time assurance claim is made, the commitment is verifiable by a separately trusted verifier using signed or cryptographically committed records, independently administered storage, or an equivalent control. The decision producer and package author are identified separately from the verifier. The verifier's administrative/signing domain or key identity is declared, and the package carries a verifier-generated signature or attestation reference evidencing participation. If the verifier shares administrative control with the producer or package author, the claim is treated as non-independent assurance and the residual common-control/failure limitation is disclosed. Where independence is not claimed, the verifier relationship and residual limitation are mandatory disclosures.

3. The evaluator availability status is recorded: callable/re-instantiable; retained under an escrow or preservation arrangement; or unavailable after a stated date. A replacement model is not presumed evidentially equivalent.
4. For any counterfactual test intended to carry later evidentiary weight, the perturbation rule, output classification rule, repetition design, semantic comparison rule (if reasons are tested), and statistical analysis are specified before the test results are examined.
5. Raw responses and relevant execution metadata are retained. Summaries alone are insufficient for independent recomputation.

### 5.3 Baseline stability calibration

Before interpreting a counterfactual difference, measure baseline instability on the same evaluator and a clearly defined calibration unit under the same declared conditions. The repetition count should not be a universal fixed number; it should be chosen ex ante from the precision or power needed for the intended claim. For illustration only, if one pre-specified calibration unit can reasonably be treated as independent Bernoulli repetitions and zero departures are observed, at least 157 repetitions are required for the upper bound of a two-sided 95% Wilson interval to fall below 2.4%. Thirty zero-departure repetitions would still have an upper Wilson bound of about 11.35%. This calculation is an illustration of precision under a simple binomial model, not a universal sample-size prescription and not a justification for pooling heterogeneous cases or evaluators.

The calibration design must therefore declare its dependence assumptions. Hosted-model backends may change, route traffic differently, cache work, or exhibit correlated behavior; different cases may also have different intrinsic instability. When multiple cases, evaluators, or repeated runs are aggregated, the analysis should preserve cluster/block identifiers and use a stratified, cluster-aware, hierarchical, or otherwise justified method rather than silently treating every response as i.i.d. The preservation package should record execution time, provider/model identifier, sampling parameters, available backend/version fingerprints, tool state, pairing/block identifiers, and other metadata needed to assess whether repetitions were made under materially comparable conditions.

### 5.4 Paired counterfactual design

Where feasible, use a paired design. For each experimental block, query the same evaluator under the unperturbed and perturbed input while holding declared execution conditions constant and preserve the pairing identifier. For one binary paired outcome per independent unit, a McNemar-type analysis or paired-proportion confidence interval of the class described by Newcombe is appropriate; non-overlap of two separately calculated Wilson intervals is not the preferred primary inference. If multiple stochastic repetitions occur within the same case/evaluator, the repeated/clustered structure must be incorporated or the results must be reported at a justified unit of analysis. For multiclass or ordinal outputs, the protocol must pre-specify an auditable transition or distance metric and a method appropriate to that outcome. No single test is asserted as universally valid across decision types.

### 5.5 Proposed decision rule

A counterfactual finding meets this protocol's quantitative admission rule only when all three conditions hold:

1. The pre-specified paired analysis yields a confidence interval for the decision difference that excludes zero (or satisfies a pre-specified equivalent hypothesis test).
2. The lower confidence bound for the pre-specified effect measure exceeds a pre-specified practical relevance threshold delta defined on the same scale. Baseline instability may inform delta only when the baseline and counterfactual quantities are commensurable; an instability rate must not be numerically compared with a different effect metric without an explicit mapping. Statistical detectability alone is not enough if the effect is smaller than the level considered operationally meaningful.
3. The evaluator, pairing, perturbation, and output-classification procedures were fixed before inspecting the counterfactual results, and complete raw responses remain available for recomputation.

If the protocol does not separate the counterfactual signal from baseline variability, the correct conclusion is "inconclusive under the pre-specified protocol." It is not evidence that no effect exists. Conversely, a large point estimate without adequate calibration does not establish causation.

### 5.6 A stricter rule for stated reasons

This paper distinguishes a model-generated natural-language justification from an EG authorization-controller reason-code (including a profile field named reason_code where such a field is used). A controller reason-code may be a deterministic policy/controller artifact bound to a final authorization record; the empirical instability reported here concerns generated explanation text and must not be generalized to deterministic controller codes without evidence. Where model-generated explanation text is offered as evidence of the grounds for authorization, that explanation channel must be calibrated and typed separately.

Unless the generated-explanation channel satisfies its own pre-specified stability and provenance test, the text should be preserved as a model-generated assertion associated with the event, not promoted to independently established evidence of why the authorization controller reached its outcome. Deterministic EG controller reason-codes are separate governance artifacts and should be analysed against their own provenance and controller logic.

### 5.7 Verification-Time Preservation Package (VTPP)

To make the preceding procedure operational, the authors propose a Verification-Time Preservation Package (VTPP) as an optional higher-assurance verification-time evidence profile. VTPP is subordinate to the EG Core Formula: it does not add a seventh live condition, create execution authority, or convert Review/Hold/Block into Proceed. It should contain, as applicable:

- Decision-State Commitment, including exact target effect and authorization outcome.
- Evaluator identity and execution manifest: provider, model/version identifier, date/time, sampling parameters, prompt/instruction stack where available, tool configuration, and provider-reported backend fingerprint where available.
- Policy, mandate, constraint, and evidence/context versions relied upon at authorization time.
- Integrity and provenance manifest using collision-resistant digests (for example SHA-256 or stronger) and signatures or equivalent controls where appropriate. The manifest identifies the decision producer and package author and their administrative domains. The digest scope and deterministic serialization/canonicalization method must be declared so that the digest is neither self-referential nor ambiguous. Repository-supplied legacy checksums such as MD5 may be retained as source-file identity metadata, but they are not treated as collision-resistant VTPP integrity proof.

- Baseline stability-calibration protocol, raw runs, and computed intervals.
- Pre-specified counterfactual protocol and raw paired responses for audit axes known at authorization time.
- Evaluator persistence status: retained/re-instantiable, escrowed, open-weight reproducible, scheduled for retirement, unavailable, or unknown. Where the package asserts reinstantiability, escrow, or open-weight reproducibility, it should identify the preservation artifact/arrangement and its limits rather than relying on the status label alone.
- Claim class for each later assurance statement: record integrity, decision-basis reconstruction, behavioral counterfactual, or generated-explanation/grounds claim.
- Verifier identity/class, administrative domain, method, and whether verifier independence is actually claimed. An independent claim requires a verifier signature or attestation reference plus a declared verifier signing domain or key identity; the validation procedure must establish that the relevant trust/control domain is separate from the decision producer/package author. Shared administrative control is labelled non-independent assurance. Where independence is not claimed, residual_limitation is required.
- Reason-channel type: model-generated natural language, structured model-generated reason, or deterministic authorization-controller reason_code.
- A statement of untested questions and any counterfactuals that cannot be answered if the evaluator later becomes unavailable.

### 5.8 Procedural Conformance Checks Beyond JSON Schema

VTPP v0.4 uses JSON Schema Draft 2020-12 to enforce field structure, conditional presence, types, patterns, and selected format syntax when format checking is enabled. Schema validity is necessary but not sufficient for a verification-time assurance claim. The standard Draft 2020-12 vocabularies do not portably express arbitrary comparisons between sibling values, trust/control relationships, or temporal inequalities; nor can a schema establish that an external signature, attestation, URL, key, or organizational relationship is genuine merely because a syntactically valid reference is present. A VTPP implementation must therefore apply semantic conformance checks in addition to schema validation. Appendix C enumerates the minimum checks proposed by this paper.

- Trust/control separation. If independence_claimed=true, the verifier administrative/control domain must resolve to a domain administratively distinct from both the decision producer and the package author. A different string is not, by itself, proof of distinct control. The declared trust_separation_class must be consistent with the resolved control relationship.
- Verifier participation. At least one verifier-generated signature or attestation reference must resolve to evidence cryptographically attributable to the declared verifier. A non-null reference alone is not evidence that the verifier handled the package or verification result.
- Signing/attestation principal control. For an independent-assurance claim, the verifier signing or attestation principal and its controlling domain must not be controlled by the decision producer or package author. verifier_signing_key_id and package_author_identity are different identifier classes and are not treated as directly comparable merely as strings.
- Temporal ordering. decision_state_commitment.commitment_created_at must be no later than authorization_event.decision_time, and verification.verified_at must be no earlier than authorization_event.decision_time. If an implementation records a distinct governed-effect commitment timestamp, the required authorization-state binding must precede that effect commitment.

- Digest and canonicalization recomputation. Collision-resistant digests must be recomputed over the bytes or objects defined by digest_scope under the declared canonicalization_method; the recomputed values must match the package and referenced artifacts.
- Claim-object consistency. Behavioral-counterfactual and generated-explanation/grounds claims must resolve to their required calibration/test objects, pairing or block identifiers, raw-response digests, and pre-specified analysis rules, with internally consistent references.
- Failure semantics. A schema-valid package that fails a required semantic conformance check must not receive an independent-assurance conformance label. The failed check and residual limitation should be preserved in the verification record.

### 5.9 Claim-Evidence Taxonomy

| Claim class | Minimum evidence basis | What it does not establish |
|---|---|---|
| Record integrity | Committed authorization/effect record; integrity/provenance verification; record identity and time | Does not reproduce model behavior or prove the underlying evidence true. |
| Decision-basis reconstruction | Record integrity plus policy/mandate, constraints, context/evidence snapshot, evaluator/version identity, and decision-controller path | Does not answer an untested behavioral counterfactual. |
| Behavioral counterfactual | Re-instantiable evaluator under declared comparable conditions, or contemporaneously preserved paired counterfactual evidence; pre-specified protocol; stability calibration; raw responses | Does not prove legal causation or generalize beyond the tested perturbation/domain. |
| Generated explanation / grounds | Clearly typed explanation channel, provenance, pre-specified semantic criterion, and channel-specific stability calibration; deterministic controller reason_codes treated separately | Generated prose is not independent testimony of internal causal reasoning. |

## 6. Worked Interpretation of the Reprocessed and Post-Hoc Measurements

The original experiment establishes two within-family behavioural comparisons, not provider succession. Llama 3.1 8B versus Llama 3.3 70B reverses 26/50 decisions (52%), but the 98% approval rate of the first model makes it a near-degenerate boundary comparison. GPT-OSS 20B versus GPT-OSS 120B reverses 15/50 (30%) in both directions, but OpenAI released the two models together; it is a sibling/model-scale comparison rather than a succession sequence. The corrected baseline v1.1 therefore preserves the measurements while correcting the relationship label.

The provider-designated replacement-path reanalysis asks a different, post-hoc question. Groq's actual migration map pairs Llama 3.3 70B with GPT-OSS 120B and Llama 3.1 8B with GPT-OSS 20B. Re-pairing the released outputs gives 19/50 reversals (38%) on the former and 32/50 (64%) on the latter. The 38% path is the stronger non-degenerate observation; the 64% path remains a boundary case because the starting model approves 49/50 cases. Neither figure should be presented as if it came from a pre-specified controlled replacement-path experiment, because the cross-family calls used different max_tokens settings and GPT-OSS alone used reasoning_effort=low.

Within-version variability remains a separate calibration issue. In each original family, 2 departures occurred in 300 repeated responses, a descriptive response-level rate of 0.67%, and 2 of 100 model-case pairs exhibited decision variation. The simple-binomial 95% Wilson interval for 2/300 is 0.18-2.40%, but it does not establish an inferential noise floor because the observations are clustered within model-case pairs and share hosted-backend conditions. The result shows that non-zero within-version variability existed in the tested setting and therefore must be measured for any counterfactual claim; it is not a universal constant.

The generated-reason channel is more variable than the decision label under an exact-text criterion, but the claim must stay descriptive. In the original within-family data, Family I has 40/100 model-case pairs with exact reason-text variation, of which 38/100 retain unanimous decisions; Family II has 24/100 reason-varying pairs, of which 22/100 retain unanimous decisions. Mean within-family reason-token Jaccard is 0.257 and 0.342. Under the post-hoc provider-designated replacement paths, mean lexical overlap is 0.2783 and 0.2610. These tokenizer/normalization-specific values are not semantic-equivalence tests, do not establish a universal explanation-instability rate, and do not justify the earlier 36-fold multiplier.

## 7. Limitations and Non-Claims

- This manuscript now claims independent raw-file reprocessing of the released Study 2 artifacts, with file identity checked against Zenodo-displayed MD5 metadata and new SHA-256 values recorded in the companion evidence manifest. It does not claim a fresh rerun of the 600 hosted-model calls. Provider-side/model-execution reproduction would be a stronger but separate experiment.
- The 22-event retirement-census arithmetic is independently reproducible from the frozen v1.1 CSV: median 16.45 months, mean 18.72, range 3.9-40.3, and 17/22 below 24 months. Exact-byte identity of the directly supplied CSV is independently confirmed by matching SHA-256 b63a15f5d87ae87b7b9b5afc0fced7acac8d50e344ebd2e998fa76a76ff9d324 and MD5 9a14876a3e48a5283cfffdc190f5f329. The verification pass establishes retirement-date provenance and the stated channel corrections; it does not establish census completeness or redefine every historical release_date as the first callable date on the verified retirement surface. The two Gemini 2.5 rows remain frozen historical planned-date observations rather than current fixed retirement dates.
- The original 50-case experiment does not establish a universal replacement or succession reversal rate. Its two specified comparisons are within-family behavioural comparisons. The provider-designated replacement-path figures (38% and 64%) are post-hoc reanalyses of already released outputs. They are informative precisely because they map to Groq's published migration paths, but they inherit a cross-family invocation asymmetry: Llama used max_tokens=120 without a reasoning-effort parameter, whereas GPT-OSS used max_tokens=900 and reasoning_effort=low. The 38% path is the stronger non-degenerate observation; the 64% path remains a boundary observation because the starting model approves 49/50 cases.
- The repeated-run experiment uses three repetitions per model-case pair. The 0.67% response-level rate and its simple Wilson interval are descriptive; clustering, case heterogeneity, and shared backend conditions prevent this manuscript from treating 2/300 as a universal i.i.d. noise-floor estimate.
- The exact-text reason-variation figure is criterion-dependent. Semantic or structured-reason analyses may produce materially different results.
- The proposed statistical protocol requires the chosen unit of analysis and dependence structure to match the data-generating process. Backend changes, caching, routing, model updates, repeated calls within the same case, and cross-call dependence must be considered. Pooled response counts must not be treated as independent merely because they are numerous.

- The protocol is not a rule of legal evidence and does not determine what a court, regulator, or tribunal must admit. Jurisdiction-specific admissibility, discovery, confidentiality, privacy, and trade-secret rules remain outside scope.
- Preservation cannot answer unknown future counterfactuals if the original evaluator cannot be re-instantiated and the relevant perturbations were never tested. VTPP makes this limitation explicit rather than hiding it behind a replacement model. A provider-designated replacement can be operationally useful without being evidentially equivalent to the retired evaluator.

## 8. Implications for Execution Governance and AI Assurance

The main governance consequence is that post-hoc reviewability is partly determined before the effect occurs. If a consequential authorization relies materially on a model that may later be retired, a governance system that records only the final decision preserves too little. A system that preserves the exact authorization state but not the evaluator preserves what was relied upon, but may still lose open-ended behavioral auditability. A system that preserves both a verifiable decision state and either an executable evaluator or contemporaneously generated counterfactual evidence has a materially stronger basis for later review.

This strengthens rather than replaces EG 3.0. Commitment-time authority remains governed by the existing six-condition Core Formula, Authorization Continuity, RecordBound and CommitCoupled semantics. The verification-time constructs in this paper are downstream assurance specializations: they describe what evidence remains examinable at t_v after an EG-governed effect has already crossed the commitment boundary. They do not create authority, alter a prior outcome, or add a seventh live condition.

The distinction also matters for assurance claims. "The record is intact" is not equivalent to "the decision basis is reconstructable"; neither is equivalent to "the evaluator can be behaviorally interrogated under an unseen counterfactual." Likewise, "the replacement model gives the same answer" is not equivalent to "the retired evaluator would have done so." The proposed VTA predicate and VTPP taxonomy force an assurance case to state which of these claims it actually supports, on which service surface, and which claims have become unavailable.

## 9. Conclusion

Model-mediated authorization creates a verification-time dependency that ordinary record retention does not resolve. Records can preserve the authorization state; separately trusted verification can substantiate the integrity and provenance of that state for an independent-assurance claim; but counterfactual auditability additionally depends on evaluator availability or on evidence deliberately preserved while the evaluator remains available.

The empirical takeaway is therefore not the largest headline reversal percentage. The original within-family measurements remain valid after their relationship label is corrected. The provider-designated post-hoc reanalysis adds a more deployment-relevant observation: the non-degenerate Llama 3.3 70B -> GPT-OSS 120B path reverses 38% of the tested modal decisions, while the 64% Llama 3.1 8B -> GPT-OSS 20B path is a boundary case with a 98% starting approval rate. Because the cross-family invocation parameters were not symmetric, these figures motivate direct replacement validation; they do not estimate a universal causal replacement effect.

The operational response is therefore not to treat replacement substitution as an evidentiary default and not to select a universal repetition count after seeing the data. It is to preserve the authorization state, record the evaluator together with its hosting/service surface, declare evaluator availability, pre-specify the calibration unit and dependence assumptions, use paired or cluster-aware counterfactual methods appropriate to the data-generating process, calibrate generated reasons separately, and retain the complete evidence needed for later recomputation. Where these conditions are absent, the governance system should say so explicitly. The strongest claim is not that every past decision can be reconstructed; it is that the boundary of what can and cannot be verified was itself governed and recorded before that boundary disappeared.

## 10. Contribution and Data Availability

Contribution statement. KU contributed the canonical EG 3.0 architectural baseline, verification-time derivation and crosswalk, claim-boundary corrections, independent Study 2 reprocessing, the external provenance check that identified the unsupported succession label, and the provider-designated replacement-path post-hoc reanalysis. Siddiqui contributed the empirical retirement census, the original within-family and repeated-run experiments, released instruments and data, the initial six-page draft, the v1.1 baseline correction/erratum, the 22-row provider-source verification pass, and the identification of the Family I near-degenerate-baseline limitation and cross-field independence-conformance gap. The verification-time assurance predicate, claim-evidence taxonomy, VTPP profile, stability-calibration logic, paired counterfactual design, corrected empirical relationship taxonomy, procedural conformance checks, and separate treatment of the generated-explanation channel are jointly developed/interpreted in this manuscript.

Data availability. The corrected baseline is The Retiring Witness v1.1, DOI: 10.5281/zenodo.22122318, which adds the erratum, retirement_census_v1_1_verified.csv, and README_retirement_census_v1_1.txt to the prior released artifacts. Independent Study 2 reprocessing used the original raw files and the R1 SHA-256 manifest described in this paper. The exact frozen census CSV supplied directly by the coauthor on 28 August 2026 independently hashes to SHA-256 b63a15f5d87ae87b7b9b5afc0fced7acac8d50e344ebd2e998fa76a76ff9d324 and MD5 9a14876a3e48a5283cfffdc190f5f329, matching the frozen README digest and the repository MD5 recorded for the census. The directly supplied README hashes to SHA-256 4a051d29213aa83c9136e627b841c65027125a17e4393cdb3957b0f47b2230a2. The README companion-dataset header still names the earlier v1.0 DOI 10.5281/zenodo.21737306; this is treated as a version-reference ambiguity rather than a data change, and this joint paper cites the current v1.1 DOI.

## Appendix A. Claim-Verification and Correction Ledger

| Issue | Finding | Revision made |
|---|---|---|
| EU AI Act horizon | Ten-year rule applies to specified Article 18 documentation; logs have a separate minimum-six-month rule under Article 19. | Corrected abstract, Sections 1 and 3.1; removed claim that decisions are legally examinable for a universal 120 months. |
| 30-repetition floor | Statistically unsupported. With 0/30 departures, 95% Wilson upper bound is about 11.35%, not below 2.4%. | Replaced fixed floor with pre-specified precision/power design; 157 zero-departure runs shown only as an illustration for a 2.4% Wilson upper bound. |
| 36x justification ratio | Mixed pair-level and response-level denominators; raw-file reprocessing also shows 24 GPT-OSS reason-varying pairs but only 22 with unanimous decisions. | Removed ratio; corrected unanimity wording; added family-specific 40/38 (Llama) and 24/22 (GPT-OSS) exact-text counts. |
| Non-overlap of separate Wilson intervals | Conservative heuristic, not the preferred inferential method for paired data. | Replaced with paired-binomial analysis and a separate practical-relevance/noise tolerance. |
| Successor as inadmissible proxy | Too categorical for two tested model families / one domain. | Reframed: equivalence cannot be presumed; direct validation is required. |
| Original relationship taxonomy and 52% / 30% results | Raw measurements reproduce, but the original predecessor-successor label was not established by the Study 2 design. GPT-OSS 20B/120B were released together; Groq published different replacement paths. | Version 8.3 retains 52% / 30% as original within-family behavioural comparisons and cites The Retiring Witness v1.1 erratum correcting the relationship taxonomy. |
| MD5 described as proof | MD5 is unsuitable as a modern collision-resistant provenance control. | VTPP specifies collision-resistant hashes such as SHA-256 plus signatures/equivalent provenance controls where appropriate. |
| Legal admissibility language | Jurisdiction-specific and unsupported by this technical note. | Defined "admission" as the paper's proposed audit/evidentiary gate, not a legal conclusion. |

| Issue | Finding | Revision made |
|---|---|---|
| EG terminology alignment | The exact labels Decision-State Commitment, Independent Verifiability, and Counterfactual Auditability are not treated as canonical EG 3.0 Core Formula primitives. | Reframed as verification-time specializations derived from canonical RecordBound/EAR/Governed Effect Record, Reviewability, Sufficient Verifiable Proof, verifier-independence, and Material Evolution semantics. |
| Independent-verifier requirement | Independent assurance needed evidence of verifier participation and a categorical common-control distinction. | Version 8.3 / VTPP v0.4 require producer/package-author identities, verifier domain/key, verifier signature/attestation, classify shared administrative control as non-independent, and add procedural checks for control-domain separation and attributable verifier participation. |
| Generated reason vs EG reason_code | v2 could be read as applying model-generated explanation instability to deterministic authorization-controller reason codes. | Separated the channels explicitly and confined the empirical instability claim to the tested model-generated explanation text. |
| VTPP status | Could be misread as an additional EG requirement. | Retained as optional non-authorizing higher-assurance profile; machine-readable companion hardened through v0.4. |
| Google lifecycle wording | Google's general lifecycle page publishes retirement schedules/lists; a separate Google page for open MaaS models explicitly states permanent endpoint deactivation and failed API requests at retirement. Anthropic states retired models are unavailable on its operated platforms and requests fail, while partner-platform schedules may differ and weights are preserved long-term. | Scoped the managed-service claim by provider/product class and retained the distinction between endpoint availability and destruction of model weights or preservation artifacts. |
| VTA recorded-claim semantics | A recorded assertion is not by itself a sufficient preserved basis for a verification claim. | Replaced RecordedClaim(q,r) with RecordSufficient(q,r), defined by claim class and a pre-specified verification rule. |

| Issue | Finding | Revision made |
|---|---|---|
| VTPP v0.1-v0.3 JSON constraints | Independent 28 August revalidation recorded for v0.4 confirms that the schema is well formed under Draft 2020-12, the supplied example and positive claim-class controls validate under explicit format checking, and targeted negative tests enforce the intended conditional-presence rules. Cross-field organizational/control truth, cryptographic participation, external-byte digest equality and chronology remain outside portable schema validation. | Retained v0.4 and the recorded schema/example/R1-manifest SHA-256 identities in Appendix B; positive and negative test coverage is separated from the semantic/control checks in Section 5.8 and Appendix C rather than treating structural schema validity as external assurance truth. |
| Pooled response-level Wilson interval | 2/300 is arithmetically 0.67% and the simple Wilson interval is correct under an i.i.d. binomial model, but the reported responses are clustered within model-case pairs and may be heterogeneous. | Relabelled the interval as descriptive; prohibited use as a universal inferential noise floor; calibration protocol now requires cluster/block preservation and a justified dependence model. |
| Baseline tolerance metric | A counterfactual effect interval and a baseline-instability rate cannot be compared numerically unless both are defined on a commensurable scale. | Decision rule now requires a pre-specified practical threshold delta on the same effect scale; baseline instability may inform delta only through an explicit mapping. |
| VTPP package digest scope | A package_digest field inside the same manifest can be ambiguous or self-referential unless the hashed bytes and serialization are defined. | v0.3 requires digest_scope and canonicalization_method and supports either an external evidence archive or a manifest hash explicitly excluding package_digest. |
| VTPP schema identifier | v0.2 used a network URL as $id even though no public VTPP endpoint is asserted in this manuscript. JSON Schema defines $id as a URI identifier, not necessarily a network locator. | v0.4 retains an absolute URN $id, avoiding any implication that the draft research schema is already hosted at a canonical public endpoint. |
| Original Study 2 Zenodo record/version | The original Study 2 package used for independent reprocessing is version 1.0 at DOI 10.5281/zenodo.21737306 (not 21737305). The corrected baseline is now version 1.1 at DOI 10.5281/zenodo.22122318. | Use 10.5281/zenodo.21737306 only when identifying the original raw package used for reprocessing; cite 10.5281/zenodo.22122318 for the corrected baseline, erratum, and retirement census. |

| Issue | Finding | Revision made |
|---|---|---|
| Retirement census reproducibility and byte identity | Version 1.1 deposits a 22-row source-verified census/README. The arithmetic remains unchanged; channel-specific retirement dates and two soft Gemini planned-date entries are explicit. The README header still names the earlier v1.0 DOI. | Version 8.3 independently recomputes all 22 row intervals and the headline statistics. Exact frozen CSV bytes supplied directly by the coauthor match SHA-256 b63a15f5d87ae87b7b9b5afc0fced7acac8d50e344ebd2e998fa76a76ff9d324 and MD5 9a14876a3e48a5283cfffdc190f5f329; the README attachment SHA-256 is 4a051d29213aa83c9136e627b841c65027125a17e4393cdb3957b0f47b2230a2. The v1.0 DOI in the README is retained as a version-reference ambiguity, not a data change. |
| Within-family and replacement-path evidentiary weight | Llama within-family 52% is near-degenerate (49/50 first-model approvals). Provider-designated post-hoc paths yield 64% (boundary) and 38% (stronger non-degenerate), but cross-family invocation parameters differ. | Version 8.3 separates original within-family comparisons from post-hoc provider-designated replacement-path reanalysis; no universal replacement effect is claimed. |
| Draft 2020-12 cross-field semantics | Standard Draft 2020-12 schema validation can enforce structure and conditional presence but does not portably prove arbitrary sibling-value inequality, organizational control separation, external signature truth, or temporal ordering. Format is annotation by default unless assertion behavior is explicitly enabled. | Added Section 5.8 and Appendix C. Independent-assurance conformance now requires application-level trust/control, signature/attestation, chronology, digest and claim-object consistency checks in addition to schema validation. |
| Commitment-time terminology | The v0.4 field commitment_created_at is the creation time of the Decision-State Commitment, not the later commitment of the governed effect. The schema description already requires it no later than the authorization decision. | Version 8.3 states the exact invariant commitment_created_at <= decision_time and separately requires verified_at >= decision_time; any distinct governed-effect commit timestamp, if implemented, is treated as a separate event that must occur after the required authorization-state binding. |

| Issue | Finding | Revision made |
| --- | --- | --- |
| Verifier key vs actor identity | verifier_signing_key_id and package_author_identity belong to different identifier classes; direct raw-string inequality is not a meaningful substitute for ownership/control analysis. | Version 8.3 requires the verifier signing/attestation principal and its controlling trust domain to be independent of the producer/package author when independence is claimed. Direct key-to-key comparison would require compatible key-identity fields or a resolver, not heterogeneous-string comparison. |
| Provider-designated replacement-path provenance | Groq explicitly recommends Llama 3.1 8B -> GPT-OSS 20B and Llama 3.3 70B -> GPT-OSS 120B or Qwen. The cross-GPT paths can be re-paired from released outputs without new calls. | Added Section 3.4 and Table 1. Results 64% / 38% are labelled post-hoc, with 38% the stronger non-degenerate observation. |
| Cross-family invocation asymmetry | run.php used max_tokens=120 and no reasoning_effort for Llama; max_tokens=900 and reasoning_effort=low for GPT-OSS. | Disclosed in abstract, Section 3.4, worked interpretation and limitations; prevents presenting replacement-path reanalysis as a pre-specified controlled head-to-head experiment. |
| Channel-specific retirement | Anthropic-operated and Google/Vertex surfaces can retire the same model version on different dates; v1.1 corrects two overbroad channel labels. | Version 8.3 states that evaluator availability identity includes hosting/service surface and time, not model ID alone. |
| Gemini 2.5 planned-date drift | Google's 2 April 2026 release notes recorded 16 October 2026; the current lifecycle table lists 20 October 2026. The frozen v1.1 rows were already flagged soft_no_earlier_than. | Version 8.3 treats 16 October only as the frozen historical planned-date observation used by the baseline and does not present it as the current fixed retirement date. |

| Issue | Finding | Revision made |
|---|---|---|
| Zenodo v1.1 public metadata consistency | The v1.1 Version Note/erratum corrected the Study 2 relationship taxonomy. In 28 August 2026 coauthor correspondence, Siddiqui confirmed that the public Description was also updated to remove successor-version framing and the overbroad 120-month evidentiary-horizon / 7.3-fold-shortfall characterization. | The manuscript no longer carries the former pre-release metadata gate. The public paper uses the corrected within-family/replacement-path taxonomy and distinguishes Article 18 documentation retention from Article 19 log retention. The metadata-closure statement is based on the coauthor confirmation; it is not represented here as an independently live-scraped Zenodo check. |

## Appendix B. VTPP v0.4 Machine-Readable Profile Verification

The associated release evidence records an independent 28 August 2026 revalidation of the companion Draft 2020-12 JSON Schema v0.4 and conforming example. The recorded result is that the schema is well formed under Draft 2020-12 and the example validates with zero errors when format checking is explicitly enabled. Recorded SHA-256 identities are: VTPP_Draft_Schema_v0.4_2026-08-25.json = b1d52c6c91b006a2a57b10715a6db3931791af6fdd600cb207e98af7939376d3; VTPP_Example_v0.4_2026-08-25.json = ee5ea324dda735c188ea5001986100a464c57385fcf0461988534f9f8328f678; and R1_Independent_Reprocessing_Manifest_v1.0_2026-08-25.json = da43ef5d67241153bc9b53a287cfc151808c1853cbe9ce2d072eeb7557bb998f. Positive control instances for behavioural_counterfactual and generated_explanation_grounds validate when their required objects are supplied, and targeted negative tests exercise the intended conditional-presence rules. Standard Draft 2020-12 schema validity is not treated as proof of organizational independence, external signature truth, digest equality, or temporal ordering; those remain procedural conformance checks under Section 5.8 and Appendix C. The $id remains an absolute URN rather than an asserted network endpoint.

| Field group | Purpose |
|---|---|
| profile / schema identity | Identifies VTPP schema version and the implementation profile; explicitly confirms that the package does not alter the EG Core Formula or create authority. |
| authorization_event | Binds record ID, decision time/outcome, exact governed-effect reference/hash plus its declared canonicalization method, authorization-record reference, commitment reference, and Governed Effect Record reference where applicable. |
| evaluator_manifest | Records provider/model/version, execution time, sampling parameters, backend fingerprints where available, availability check time, preservation/re-instantiation status, expected retirement date where asserted, preservation reference where required, and comparability limits. |
| decision_state_commitment | Binds policy/mandate, constraints, context/evidence, input/prompt/instruction/tool manifests, commitment creation time, and collision-resistant commitment digest. |

| Field group | Purpose |
|---|---|
| provenance | Identifies decision producer and package author plus their administrative domains; requires collision-resistant package digest, digest scope/canonicalization, custody and package-author signature/attestation where used; legacy MD5 may be source metadata only. |
| verification | Declares claim class, verifier identity/class, administrative/signing domain or key, trust-separation class, method, time, result and limitations. Independent claims require verifier-generated signature/attestation evidence; non-independent claims require residual_limitation. Cross-field trust/control separation and temporal ordering are checked procedurally under Section 5.8 / Appendix C rather than inferred from schema validity alone. |
| stability_calibration | Records protocol registration time, repetitions, metric, confidence method/level, raw-response digest, estimate and interval; required for behavioral-counterfactual and generated-explanation/grounds claims, with dependence assumptions declared by the protocol. |
| counterfactual_test | Records pairing, perturbation specification, analysis method, practical threshold, raw paired-response digest, and result; required for behavioral-counterfactual claims. |
| reason_channel | Separates model-generated natural-language explanation, structured model-generated reason, and deterministic authorization-controller reason-code; generated-explanation claims require a pre-specified semantic rule and calibration reference. |
| limitations | Records untested questions, unavailable counterfactuals, external assumptions, known gaps, and any comparability or retention limits. |

## Appendix C. VTPP v0.4 Procedural Conformance Checklist

This appendix defines semantic checks that supplement, rather than replace, Draft 2020-12 schema validation. A package can be structurally schema-valid while still making an unsupported independence, chronology, provenance, or artifact-integrity claim. The proposed conformance decision therefore evaluates both schema assertions and relationships that require application-level evidence or cross-field comparison.

- **C1 - Schema and format validation.** Validate the package against the declared VTPP v0.4 Draft 2020-12 schema. Where date-time/URI format semantics are relied upon, the validator must explicitly enable or otherwise implement the intended format checks; format annotation alone is not treated as external truth.
- **C2 - Actor and domain resolution.** Resolve decision_producer_identity, package_author_identity and verifier_identity to the principals actually responsible for those roles, and resolve their declared administrative/control domains. Unresolvable identities or domains fail the corresponding assurance claim.
- **C3 - Independence/control separation.** If independence_claimed=true, the verifier must be administratively/control-separated from both the decision producer and package author. Canonicalized domain labels should differ, but string inequality is not sufficient; available governance, key-control, employment, delegation, or custody evidence must not contradict the declared administratively_separate trust class.

- **C4 - Verifier participation evidence.** Resolve verifier_signature_ref or verifier_attestation_ref and verify that the evidence is cryptographically attributable to the declared verifier. A syntactically valid reference, by itself, does not substantiate verifier participation.
- **C5 - Signing/attestation principal control.** Where independent assurance is claimed, the verifier signing/attestation principal and the controlling key or trust domain must not be controlled by the decision producer or package author. verifier_signing_key_id and package_author_identity are different identifier classes and should not be compared as raw strings as a substitute for control analysis.
- **C6 - Temporal ordering.** Check decision_state_commitment.commitment_created_at <= authorization_event.decision_time and verification.verified_at >= authorization_event.decision_time. The first timestamp is the creation of the Decision-State Commitment, not the later commitment of the governed effect. Where an implementation records a distinct governed-effect commitment timestamp, the required authorization-state binding must precede that effect commitment.
- **C7 - Integrity and canonicalization.** Recompute each required collision-resistant digest over the declared digest_scope using the declared canonicalization_method, and confirm that the referenced bytes/objects and recomputed digest agree. Legacy repository MD5 values may be retained only as source metadata, not as collision-resistant VTPP integrity evidence.
- **C8 - Claim-specific object consistency.** For behavioral-counterfactual claims, resolve stability_calibration and counterfactual_test objects, pairing/block identifiers, raw-response digests and pre-specified analysis rules. For generated-explanation/grounds claims, resolve the reason-channel type, semantic comparison rule and channel-specific calibration. Missing or internally inconsistent objects fail the corresponding claim class.